\documentclass[conference]{IEEEtran}
\IEEEoverridecommandlockouts
\usepackage{cite}
\usepackage{amsmath,amssymb,amsfonts}
\usepackage{algorithm}
\usepackage{algorithmic}
\usepackage{graphicx}
\usepackage{textcomp}
\usepackage{xcolor}
\usepackage{multirow}
\usepackage{url}
\def\BibTeX{{\rm B\kern-.05em{\sc i\kern-.025em b}\kern-.08em
    T\kern-.1667em\lower.7ex\hbox{E}\kern-.125emX}}
\begin{document}

\title{Verifying Integrity of Deep Ensemble Models by Lossless Black-box Watermarking with Sensitive Samples}

\author{Lina Lin and Hanzhou Wu\\
Shanghai University\\
\thanks{This work was supported in part by the National Natural Science Foundation of China under Grant number 61902235, and in part by the Shanghai ``Chen Guang'' Program under Grant number 19CG46.}
\thanks{\emph{Lina Lin is currently an undergraduate student and will pursue her master degree under supervision of Dr. Hanzhou Wu started from September 2022.}}
\thanks{\emph{Corresponding author: Hanzhou Wu (contact email: h.wu.phd@ieee.org)}}
}

\maketitle

\begin{abstract}
With the widespread use of deep neural networks (DNNs) in many areas, more and more studies focus on protecting DNN models from intellectual property (IP) infringement. Many existing methods apply digital watermarking to protect the DNN models. The majority of them either embed a watermark directly into the internal network structure/parameters or insert a zero-bit watermark by fine-tuning a model to be protected with a set of so-called trigger samples. Though these methods work very well, they were designed for individual DNN models, which cannot be directly applied to deep ensemble models (DEMs) that combine multiple DNN models to make the final decision. It motivates us to propose a novel black-box watermarking method in this paper for DEMs, which can be used for verifying the integrity of DEMs. In the proposed method, a certain number of sensitive samples are carefully selected through mimicking real-world DEM attacks and analyzing the prediction results of the sub-models of the non-attacked DEM and the attacked DEM on the carefully crafted dataset. By analyzing the prediction results of the target DEM on these carefully crafted sensitive samples, we are able to verify the integrity of the target DEM. Different from many previous methods, the proposed method does not modify the original DEM to be protected, which indicates that the proposed method is lossless. Experimental results have shown that the DEM integrity can be reliably verified even if only one sub-model was attacked, which has good potential in practice. 
\end{abstract}


\section{Introduction}
\IEEEPARstart{T}{here} is no doubt that deep neural networks (DNNs) \cite{DLNature} have brought profound changes to our society because of its superior performance in many application areas such as computer vision, speech recognition, natural language processing and games. It can be foreseen that DNNs will continue to make great contributions. However, building a superior DNN model with superior performance on its original task needs a lot of computing resources, large quantities of well-labelled data and expertise of structural design \cite{Zhao:WIFS}. Moreover, due to the full openness or semi-openness of DNN models, it is very easy for a malicious user to illegally copy, download, steal, tamper and sell DNN models. It indicates that as an expensive digital asset, it is necessary to protect the intellectual property (IP) of DNN models, which promotes \emph{watermarking DNNs} (or \emph{DNN watermarking}) \cite{Uchida:2017} to become a popular topic recently. 

A straightforward method is to extend media watermarking to DNN watermarking directly, which, however, is not desirable for practice since simply modifying the network weights may significantly degrade the performance of the DNN model on its original task and thus reduces the commercial value of the DNN model. Therefore, we need to design watermarking schemes specifically for DNNs. Since around 2017, increasing DNN watermarking methods are introduced in the literature. These methods can be roughly divided to two categories, i.e., \emph{white-box DNN watermarking} and \emph{black-box DNN watermarking}. The former requires the model owner to access the target model including its structure and internal parameters to extract the embedded watermark such as \cite{Uchida:2017, Li:2021, Wang:2020, Fan:2019, Chen:2019, Wang:2021, Zhao:WIFS}. However, in practice, it is more likely that we have not the access to the internal details of the target DNN model, which motivates researchers to propose black-box DNN watermarking algorithms. The term ``black-box'' indicates that the watermark should be extracted without knowing the internal details of the DNN. By reviewing the existing black-box watermarking methods, it is often the case that the watermark is extracted by querying the target DNN model and checking the DNN outputs in correspondence to a set of carefully crafted input samples also called \emph{trigger samples}. In order to embed a secret watermark, the host DNN is generally trained or fine-tuned in such a way that the marked model outputs the expected results matching the watermark when inputting a sequence of trigger samples during ownership verification, while maintaining the performance on its original task. Many works are designed along this line such as \cite{Zhao:ISDFS, Adi:2018, Zhang:AsiaCCS, Wang:Sym2022, Wu:2021}.

As an efficient strategy to boost performance, ensemble allows us to combine multiple DNNs to make the final decision. By deploying deep ensemble models (DEMs) in commercial products, technology companies can provide the higher-quality smart services. Ensemble architecture has played an important role in large-scale intelligent service systems and would surely become more and more popular in future. It is very necessary to prevent the DEMs from modification without permission, for which watermarking is one of the most suitable solutions. However, regardless of superiority, the aforementioned DNN watermarking methods are originally designed for individual DNN models, which cannot be directly applied to ensemble systems. Moreover, it has been demonstrated that ensemble is a very effective strategy to easily evade the verification of copyright infringements from the legitimate owners \cite{Hitaj:2021} under the black-box condition, meaning that the above trigger-sample based methods may not work when to apply them to DEMs. It motivates us to investigate DEM watermarking in this paper. 

We present the design and implementation of a novel black-box watermarking method that allows the legitimate owner of the DEM to determine whether the DEM was modified before or not. The proposed method completes the verification by a remote fashion, which is very suitable for real-world scenarios. Unlike the aforementioned black-box watermarking methods that inevitably distort the original model so as to produce the watermarked model, the proposed method leaves the original DEM unchanged and therefore will not introduce noticeable distortion to the DEM and not impair the generalization ability of the host DEM. In the proposed method, a certain number of sensitive samples are selected by mimicking real-world attacks and analyzing the prediction results of the probably attacked sub-models of the original DEM on a carefully crafted dataset. By analyzing the prediction results of the target DEM on these sensitive samples, we are able to verify the integrity of the DEM. Experiments have shown that the DEM integrity can be reliably verified even if only one sub-model was attacked.

The rest structure of this paper is organized as follows. We first introduce the proposed method in Section II, followed by experiments in Section III. Finally, we conclude this paper in Section IV. 

\section{Proposed Method}
In this section, we will first formulate our problem and then generate the sensitive samples used for integrity verification.

\subsection{Problem Formulation}
We limit the original task of the DEM to be watermarked to image classification. Let $\mathcal{E}: \mathbb{R}^d \mapsto \mathcal{C}$ represent the DEM which consists of $n$ DNN models $\mathcal{M}_1$, $\mathcal{M}_2$, ..., $\mathcal{M}_n$, where $\mathcal{M}_i: \mathbb{R}^d \mapsto \mathcal{C}$ is the $i$-th DNN model, $\mathcal{C} = \{1, 2, ..., c\}$ includes all possible classification results and $c \geq 2$ is the total number of classes. The classification result of $\mathcal{E}$ on a sample $\textbf{x}\in \mathbb{R}^d$ is
\begin{equation}
\mathcal{E}(\textbf{x}) = \text{En}(\mathcal{M}_1(\textbf{x}), \mathcal{M}_2(\textbf{x}), ..., \mathcal{M}_n(\textbf{x}))\in \mathcal{C},    
\end{equation}
where $\mathcal{M}_i(\textbf{x})\in \mathcal{C}$ for $1\leq i\leq n$ and $\text{En}(\cdot,\cdot,...,\cdot)$ represents the ensemble mechanism such as majority voting. $\mathcal{E}$ is trained on an ordinary dataset $\mathcal{D}$ = $\{(\textbf{x}_i, y_i)~|~1\leq i\leq |\mathcal{D}|\}$, where $y_i\in \mathcal{C}$ is the ground-truth label of $\textbf{x}_i\in \mathbb{R}^d$ and $|\mathcal{D}|$ counts the total number of image samples in $\mathcal{D}$.

Our main purpose is to watermark $\mathcal{E}$ by constructing a set of sensitive samples so that the integrity of $\mathcal{E}$ can be verified by analyzing the prediction results of the target DEM $\mathcal{E}' \approx \mathcal{E}$ in correspondence to the constructed sensitive samples. Since we do not modify the original $\mathcal{E}$, the released DEM (i.e., $\mathcal{E}$) for practical use will not arouse any suspicion from the adversary. Moreover, the generalization ability of $\mathcal{E}$ will not be impaired. 

Mathematically, given a set of sensitive samples $S = \{\textbf{s}_i|i\in [1, |S|]\}$, $\mathcal{E}'$ is treated as a \emph{tampered} version of $\mathcal{E}$ if
\begin{equation}
\frac{1}{|S|}\sum_{i=1}^{|S|}\delta(\mathcal{E}(\textbf{s}_i), \mathcal{E}'(\textbf{s}_i)) < 1 - \epsilon,
\end{equation}
where $0\leq \epsilon< 1$ is a very small threshold, and $\delta(x,y) = 1$ if $x=y$ otherwise $\delta(x,y) = 0$. Accordingly, we are capable of verifying the integrity of the target DEM to be protected. It can be inferred that the proposed method corresponds to a \emph{zero-bit watermarking} scheme. The most important task is to construct $S$, which will be introduced in the next subsection. We argue that though $\epsilon = 0$ enables us to determine whether the target model was modified or not, it is more desirable by setting a small $\epsilon > 0$ from the viewpoint of generalization. 

\subsection{Sensitive Sample Generation}
Throughout this paper, we will use the most popular ensemble strategy \emph{majority voting}, which can be expressed as
\begin{equation}
\mathcal{E}(\textbf{x}) = \arg\max_{j} \sum_{i=1}^{n}\delta(\mathcal{M}_i(\textbf{x}), j),
\end{equation}
for which when there are multiple feasible solutions, any one among them can be \emph{randomly} selected as the final result. Majority voting makes the predictions of sub-models independent of each other so that the result of ensemble is robust to attacks. For example, the result of ensemble may be unchanged even though only one sub-model was tampered, resulting in that the integrity of the ensemble model is difficult to be verified. If the sub-models are not independent of each other, e.g., taking the average value of the softmax outputs of the trained sub-models as the result of ensemble, the ensemble model can be regarded as a single model so that conventional watermarking methods can be used. Therefore, it is very appropriate to verify the integrity of majority voting based DEM specifically. 

Based on the aforementioned analysis, we propose a method to verify the integrity of the target DEM through constructing a certain number of sensitive samples. The sensitive samples are sensitive to arbitrary modifications to the sub-models. In other words, our purpose is to ensure that even though only one sub-model was intentionally modified, the probability resulting in a different ensemble result on a single sensitive sample should be high, which can be mathematically expressed as
\begin{equation}
\text{Pr}\left [\delta(\mathcal{E}(\textbf{s}_i), \mathcal{E}'(\textbf{s}_i)) = 0 \right ] \geq 1 - \epsilon, \forall~\textbf{s}_i \in S.
\end{equation}
Obviously, when $|S|$ is large enough, Eq. (2) will surely hold if Eq. (4) holds from the point of view of statistical analysis. Our task is now to generate $S$. The details are provided below.

First, a set of candidate samples $S_c = \{\textbf{c}_1, \textbf{c}_2, ..., \textbf{c}_{|S_c|}\}$ are collected. It is required that $|S_c| >> |S|$. There are different ways to construct $S_c$. For example, each sample in $S_c$ may be randomly generated according to a secret key, i.e., each sample in $S_c$ is a random image, whose entropy is maximized. One may also collect other meaningful images related/unrelated to the original task of the target DEM for constructing $S_c$. When the images related to the original task are used, the contents of these images may need to be modified to generate the sensitive samples, e.g., by applying adversarial perturbations \cite{APSurvey}. 

Second, for each sub-model $\mathcal{M}_i$, $1\leq i\leq n$, we mimic a real-world attack to $\mathcal{M}_i$ and produce a sequence of attacked sub-models, denoted by $A_i = \{\mathcal{M}_{i,1}, \mathcal{M}_{i,2}, ..., \mathcal{M}_{i,|A_i|}\}$. For the sake of simplicity, we assume that $|A_1| = |A_2| = ... = |A_n|$. Each sub-model $\mathcal{M}_{i,j}$ may involve more than one attack. E.g., $\mathcal{M}_{i,j}$ can be produced by first compressing the sub-model $\mathcal{M}_i$ and then fine-tuning the compressed sub-model. Assuming that we have already collected all the attacked sub-models, for each candidate sample $\textbf{c}_k\in S_c$, we determine
\begin{equation}
\rho_i(\textbf{c}_k) = 1 - \frac{1}{|A_i|}\sum_{j=1}^{|A_i|}\delta(\mathcal{M}_i(\textbf{c}_k), \mathcal{M}_{i,j}(\textbf{c}_k)), \forall i\in [1, n].
\end{equation}

It can be inferred from Eq. (5) that a higher $\rho_i(\textbf{c}_k)$ indicates that $\textbf{c}_k$ is more sensitive to modifications to $\mathcal{M}_i$. We construct a \emph{binary vector} $\textbf{b}_k = (b_1, b_2, ..., b_n)\in \{0, 1\}^n$ for $\textbf{c}_k$ by
\begin{equation}
b_i = \left\{\begin{matrix}
1 & \rho_i(\textbf{c}_k) \geq \alpha,\\ 
0 & \rho_i(\textbf{c}_k) < \alpha,
\end{matrix}\right.
\end{equation}
where $0 < \alpha \leq 1$ is a pre-determined threshold. Finally, $\textbf{c}_k$ is deemed \emph{sensitive} if the Hamming weight of $\textbf{b}_k$ is no less than a threshold, namely, the percentage of ``1'' in $\textbf{b}_k$ satisfies
\begin{equation}
\sum_{j=1}^{n}\delta(b_j,1) \geq \beta n,
\end{equation}
where $0 < \beta \leq 1$ is a pre-determined threshold. In this way, we can collect a set of sensitive samples from $S_c$. The pseudo-code for sensitive sample generation is shown in Algorithm 1.

\begin{algorithm}[!t]
 \caption{Pseudocode for sensitive sample generation}
 \begin{algorithmic}[1]
	\renewcommand{\algorithmicrequire}{\textbf{Input:}}
	\renewcommand{\algorithmicensure}{\textbf{Output:}}
	\REQUIRE Sub-models $\mathcal{M}_1, \mathcal{M}_2, ..., \mathcal{M}_n$; Candidate set $S_c$; $\alpha, \beta$.
	\ENSURE  Sensitive set $S$.
	\FOR {$i$ = 1, 2, ..., $n$}
	    \STATE Mimic the real-world attack to $\mathcal{M}_i$ to produce a set of attacked sub-models $A_i = \{\mathcal{M}_{i,1}, \mathcal{M}_{i,2}, ..., \mathcal{M}_{i,|A_i|}\}$
	\ENDFOR
	\STATE Initialize $S = \emptyset$ 
	\FOR {$k$ = 1, 2, ..., $|S_c|$}
	    \STATE Determine $\rho_1(\textbf{c}_k), \rho_2(\textbf{c}_k), ..., \rho_n(\textbf{c}_k)$ with Eq. (5)
	    \STATE Determine $\textbf{b}_k$ with Eq. (6)
	    \IF {$\textbf{b}_k$ meets Eq. (7)}
	        \STATE Append $\textbf{c}_k$ to $S$
	    \ENDIF
	\ENDFOR
	\RETURN $S$
 \end{algorithmic}
\end{algorithm}

\section{Experimental Results and Analysis}
In this section, we conduct extensive experiments to demonstrate the superiority and applicability of the proposed work.

\subsection{Setup}
We use two widely used benchmark datasets MNIST\footnote{\url{http://yann.lecun.com/exdb/mnist/}} and CIFAR-10\footnote{\url{https://www.cs.toronto.edu/~kriz/cifar.html}} for experiments. The former consists of a total of 70,000 grayscale images (each with a size of $28\times 28$) in ten classes. The latter consists of a total of 60,000 color images (each with a size of $32\times 32$) in ten classes. For each dataset, we use three sub-models to build an ensemble model to perform image classification. The ensemble model applied to MNIST consists of LeNet-5 \cite{LeNet}, VGG-19 \cite{VGG} and ResNet-18 \cite{ResNet}. The ensemble model tested on CIFAR-10 consists of AlexNet \cite{AlexNet}, VGG-19 and ResNet-18. For the MNIST dataset, it is divided into five disjoint subsets. In detail, we use 48,000 images for model training, 4,000 images for model validation, 10,000 images for model testing, 4000 images for mimicking the real-world attack to sub-models, 4000 images for realizing the real-world attack to sub-models. Similarly, for CIFAR-10, we use 40,000 images for model training, 2,000 images for model validation, 10,000 images for model testing, 4000 images for mimicking the real-world attack to sub-models, 4000 images for realizing the real-world attack to sub-models. The learning rate is set to 0.001, and the batch size is set to 64. The Adam optimizer \cite{Adam} is used for model training and the open source framework PyTorch\footnote{\url{https://pytorch.org/}} is used for simulation. For simplicity, we use ``MNIST-DEM'' and ``CIFAR-10-DEM'' to denote the ensemble model applied to MNIST and CIFAR-10. 

Two common real-world attacks are evaluated in our experiments. One is \emph{model fine-tuning}, which adjusts the parameters of a trained model in order to fit with certain observations. The other is \emph{model compression and model fine-tuning}, which first reduces the size of a deep model for lightweight deployment and then adjusts the parameters of the processed model so that the performance of the model on its original task is restored. For simplicity, we use ``MF'' and ``MC+MF'' to represent the two attacks mentioned above. During the process of generating sensitive samples, for each sub-model, the number of attacked sub-models is set to 6, i.e., $|A_1| = |A_2| = ... = |A_n| = 6$ (see Line 2 in Algorithm 1). It is noted that all the attacked sub-models are using the same type of attack. For example, suppose that we are going to perform the MF attack to get 6 attacked versions of LeNet-5, the 6 attacked sub-models are generated by fine-tuning the trained LeNet-5 with the above-mentioned 4,000 images by different epochs. Similarly, when we perform the MC+MF attack, 6 random compression rates in range [0.01, 0.3] are used, e.g., 0.01 means that 1\% weights with the smallest absolute values are removed. And, for each compression rate, the number of epochs is randomly generated for fine-tuning. The process of realizing (not mimicking) the real-world attack to a sub-model is similar. In addition, $\alpha$ in Eq. (6) is set to 2/3, and $\beta$ in Eq. (7) is set to 2/3 by default.

\begin{table*}[!t]
\renewcommand{\arraystretch}{1}
\centering
\caption{Classification accuracy on the original task. Acc-MF means the classification accuracy by applying the MF attack (if any). Acc-MC+MF means the classification accuracy by applying the MC+MF attack (if any).}
\begin{tabular}{ccc|cc|ccc|cc}
\hline\hline
\multicolumn{5}{c|}{MNIST-DEM} & \multicolumn{5}{c}{CIFAR-10-DEM}\\
\hline
LeNet-5 & VGG-19 & ResNet-18 & Acc-MF & Acc-MC+MF & AlexNet & VGG-19 & ResNet-18 & Acc-MF & Acc-MC+MF\\
\hline
- & - & - & 99.40\% & 99.40\% & - & - & - & 84.21\% & 84.21\%\\
\hline
- & - & attacked & 99.50\% & 99.07\% & - & - & attacked & 84.63\% & 85.15\% \\
- & attacked & - & 99.30\% & 98.87\% & - & attacked & - & 84.60\% & 84.04\% \\
- & attacked & attacked & 99.36\% & 98.80\% & - & attacked & attacked & 85.19\% & 85.03\% \\
attacked & - & - & 99.40\% & 98.98\% & attacked & - & - & 84.43\% & 84.38\% \\
attacked & - & attacked & 99.38\% & 99.30\% & attacked & - & attacked & 83.29\% & 84.95\% \\
attacked & attacked & - & 99.28\% & 98.80\% & attacked & attacked & - & 84.85\% & 83.92\% \\
attacked & attacked & attacked & 99.32\% & 99.02\% & attacked & attacked & attacked & 85.06\% & 84.83\% \\
\hline\hline
\end{tabular}
\end{table*}

We use two different strategies to build $S_c$. The first strategy is to generate a set of fully random images as $S_c$ according to a secret key. The second strategy is to use a set of meaningful images unrelated to the original task of the model as $S_c$. For the second strategy, in our experiments, when to evaluate the proposed work on the MNIST dataset, we use a set of images randomly selected out from the CIFAR-10 dataset to construct $S_c$ and vice versa. To select a sufficient number of sensitive images, the size of $S_c$ is set to 10,000, i.e., $|S_c|$ = 10,000. We are now ready to report our results in the next subsection. 

\begin{table*}[!t]
\renewcommand{\arraystretch}{1}
\caption{The number of generated sensitive samples.}
\centering
\begin{tabular}{c|c|c|c|c|c}
\hline\hline
Model & Attack & Source of Candidate Set & $|S_c|$ & $|S|$ & $|S|/|S_c|$\\
\hline
\multirow{4}{*}{MNIST-DEM} & \multirow{2}{*}{MF} & randomly generated (RG) & 10,000 & 600 & 6.00\%\\
 & & CIFAR-10 & 10,000 & 379 & 3.79\% \\
\cline{2-6}
 & \multirow{2}{*}{MC+MF} & RG & 10,000 & 1,052 & 10.52\%\\
 & & CIFAR-10 & 10,000 & 2,129 & 21.29\% \\
\hline
\multirow{4}{*}{CIFAR-10-DEM} & \multirow{2}{*}{MF} & RG & 10,000 & 465 & 4.65\%\\
 & & MNIST & 10,000 & 1,456 & 14.56\% \\
\cline{2-6}
 & \multirow{2}{*}{MC+MF} & RG & 10,000 & 194 & 1.94\%\\
 & & MNIST & 10,000 & 2,891 & 28.91\% \\
 \hline
 \multicolumn{5}{c|}{Average value} & 11.46\%\\
\hline\hline
\end{tabular}
\end{table*}

\begin{table*}[!t]
\renewcommand{\arraystretch}{1}
\centering
\caption{Classification accuracy on the sensitive set. Acc-RG means the classification accuracy by using the randomly generated images as the candidate images. Acc-CIFAR-10 means the classification accuracy by selecting the images from CIFAR-10 as the candidate images. Acc-MNIST means the classification accuracy by selecting the images from MNIST as the candidate images.}
\begin{tabular}{c|ccc|cc|ccc|cc}
\hline\hline
\multirow{2}{*}{Attack} & 
\multicolumn{5}{c|}{MNIST-DEM} & \multicolumn{5}{c}{CIFAR-10-DEM}\\
\cline{2-11}
& LeNet-5 & VGG-19 & ResNet-18 & Acc-RG & Acc-CIFAR-10 & AlexNet & VGG-19 & ResNet-18 & Acc-RG & Acc-MNIST\\
\hline
\multirow{7}{*}{MF} & - & - & attacked & 99.50\% & 86.28\% & - & - & attacked & 88.60\% & 82.76\% \\
& - & attacked & - & 58.83\% & 58.05\% & - & attacked & - & 91.83\% & 50.82\% \\
& - & attacked & attacked & 56.00\% & 46.44\% & - & attacked & attacked & 90.97\% & 28.57\% \\
& attacked & - & - & 81.50\% & 69.66\% & attacked & - & - & 85.16\% & 67.03\% \\
& attacked & - & attacked & 81.33\% & 58.05\% & attacked & - & attacked & 73.76\% & 58.24\% \\
& attacked & attacked & - & 42.50\% & 34.56\% & attacked & attacked & - & 90.75\% & 30.91\% \\
& attacked & attacked & attacked & 44.67\% & 28.23\% & attacked & attacked & attacked & 88.17\% & 23.21\% \\
\hline
\multirow{7}{*}{MC+MF} & - & - & attacked & 99.81\% & 88.16\% & - & - & attacked & 79.38\% & 64.37\% \\
& - & attacked & - & 98.48\% & 79.33\% & - & attacked & - & 92.27\% & 65.51\% \\
& - & attacked & attacked & 98.48\% & 72.19\% & - & attacked & attacked & 68.56\% & 43.48\% \\
& attacked & - & - & 47.34\% & 70.13\% & attacked & - & - & 80.93\% & 74.78\% \\
& attacked & - & attacked & 47.34\% & 59.84\% & attacked & - & attacked & 76.80\% & 42.93\% \\
& attacked & attacked & - & 45.82\% & 51.48\% & attacked & attacked & - & 74.23\% & 45.35\% \\
& attacked & attacked & attacked & 45.82\% & 45.42\% & attacked & attacked & attacked & 88.14\% & 29.78\% \\
\hline\hline
\end{tabular}
\end{table*}

\subsection{Results and Analysis}
Given a trained ensemble model consisting of $n$ sub-models, there are $2^n$-$1$ possible strategies to attack the ensemble model, i.e., for each $k\in [1, n]$, there are $\binom{n}{k}$ different combinations to attack exactly $k$ sub-models of the ensemble model, leading to a total of $\sum_{k=1}^n\binom{n}{k} = 2^n$-$1$ different combinations. Table I shows the classification accuracy on the corresponding dataset for the original ensemble model (which is non-attacked) and all the corresponding attacked ensemble models. It can be seen that the classification accuracy after attack is not significantly degraded, indicating that the two attack methods used in this paper are reasonable and likely to appear in real scenarios.

Table II has listed the number of generated sensitive samples under different conditions. It can be observed that, averagely, around 11\% candidate samples can be finally determined as the sensitive samples, which is sufficient for integrity verification. It should be admitted that it is always free for us to determine the source of the candidate set, which is not the main interest of this short paper. Moreover, as has shown in Eq. (6) and Eq. (7), in order to collect sensitive samples, we have to use two thresholds, which are free to preset as well.

In order to verify the integrity of the target ensemble model, we feed the sensitive samples to the target ensemble model and obtain the prediction (i.e., classification) results from the target ensemble model. Then, by comparing these prediction results with the ground-truth labels (which are the prediction results of the original non-attacked ensemble model), we can verify the integrity of the target ensemble model. Along this line, Table III provides the classification accuracy values due to different experimental conditions. Here, the determination of classification accuracy can be easily derived from Eq. (2). From Table III, we can find that: first, different attacking strategies result in different classification accuracy values, which is reasonable because different sub-models have different sensitivities to the attacks and samples, e.g., it can be inferred from Table III that ResNet-18 has less sensitivity on randomly generated samples. Moreover, the integrity verification performance is affected by $\alpha$ and $\beta$, but we did not optimize the parameters. Second, most accuracy values are significantly lower than 100\%, indicating that the proposed method has the ability to reliably verify the integrity of the target ensemble model no matter how many sub-models were attacked, which has verified the applicability.

\section{Conclusion and Discussion}
The fast development of deep learning technology has made the protection of IP of deep models a hot research topic. This work was motivated by the fact that ensemble has become a popular and effective strategy to enhance the performance of deep model, implying that developing effective methods for protecting the IP of deep ensemble models is very necessary. In order to tackle with this problem, in this paper, we propose a novel method to verify the integrity of an ensemble model by analyzing the prediction results of the model in correspondence to a set of carefully constructed sensitive samples. On one hand, we do not modify the original ensemble model at any stage which will never impair the ensemble model on its original task. In other words, our method can be considered as a special kind of lossless watermarking (or called reversible watermarking, invertible embedding) method \cite{Wu:spie, Wu:TCSVT}. On the other hand, the process of integrity verification is realized under black-box condition, i.e., the ownership verifier has no access to the internal details of the target model, which is very helpful for practice. Experimental results have shown that this work is capable of reliably verifying the integrity of the target ensemble model, which shows the applicability and superiority. In this paper, for simplicity, we use either random images or images unrelated to the original task to construct the sensitive set $S$. It is admitted that one may use other methods to construct $S$, e.g., using black-box adversarial samples.

Moreover, the proposed method is not limited to ensemble models. In other words, the proposed method can be applied to individual models for integrity verification as well, which indicates that the proposed method has good universality. In addition, by mimicking malicious attacks and non-malicious attacks and fine-tuning thresholds $(\alpha, \beta)$, the proposed method may be extended to semi-fragile, which is robust against non-malicious attacks but sensitive to malicious attacks. We leave this interesting problem as the follow-up work. 

It should be noted that during the process of constructing the sensitive samples, each sensitive sample should be uniquely classified by the original DEM. Otherwise, the integrity cannot be reliably verified due to the random classification of the original DEM. To reduce the size of side information, the classification results of the original DEM can be losslessly compressed. Alternatively, one may select the sensitive samples corresponding to a specified class (based on the classification of the original DEM) to constitute the final sensitive set. Additionally, the proposed method is a probabilistic verification method, indicating that the number of sensitive samples should not be too small. Otherwise, the integrity may not be verified.


\end{document}